\documentclass{article}

\usepackage{arxiv}

\usepackage[utf8]{inputenc} 
\usepackage[T1]{fontenc}    
\usepackage{hyperref}       
\usepackage{url}            
\usepackage{booktabs}       
\usepackage{amsfonts}       
\usepackage{nicefrac}       
\usepackage{microtype}      
\usepackage{lipsum}

\usepackage{graphicx}
\usepackage{amssymb}
\usepackage{amsmath}
\usepackage{multirow}

\usepackage[ruled,vlined]{algorithm2e}
\usepackage{algorithmic}

\title{The principle of weight divergence facilitation for unsupervised pattern recognition in spiking neural networks}

\author{
  Oleg~Nikitin\\
  Computing Center \\
  Far Eastern Branch\\ 
  Russian Academy of Sciences\\
  680000, Khabarovsk, Russia\\
  \texttt{olegioner@gmail.com} \\
   \And
 Olga~Lukyanova\thanks{Corresponding author} \\
  Computing Center \\
  Far Eastern Branch\\ 
  Russian Academy of Sciences\\
  680000, Khabarovsk, Russia\\
  \texttt{ollukyan@gmail.com} \\
  \And
 Alex~Kunin \\
  Computing Center \\
  Far Eastern Branch\\ 
  Russian Academy of Sciences\\
  680000, Khabarovsk, Russia\\
  \texttt{alexkunin88@gmail.com} \\
}

\begin{document}
\maketitle

\begin{abstract}
Parallels between the signal processing tasks and biological neurons lead to an understanding of the principles of self-organized optimization of input signal recognition. In the present paper, we discuss such similarities among biological and technical systems. We propose adding the well-known STDP synaptic plasticity rule to direct the weight modification towards the state associated with the maximal difference between background noise and correlated signals. We use the principle of physically constrained weight growth as a basis for such weights' modification control. It is proposed that the existence and production of bio-chemical 'substances' needed for plasticity development restrict a biological synaptic straight modification. In this paper, the information about the noise-to-signal ratio controls such a substances' production and storage and drives the neuron's synaptic pressures towards the state with the best signal-to-noise ratio. We consider several experiments with different input signal regimes to understand the functioning of the proposed approach.
\end{abstract}



\keywords{Neural homeostasis \and Spike-timing-dependent plasticity \and Synaptic scaling \and Adaptive control \and Bio-inspired cognitive architectures \and Neural networks \and Neural network architectures}


\section{Introduction}

Signal processing is a well-known field of computer science. Since the invention of radio signal transmission, there was a task of optimal signal recovery from noise. Correlators represent the mathematical approach to solve the noise filtering task. They perform a simple mathematical operation to estimate the correlation coefficient among all input sources and may be implemented as special hardware. The fundamental goal of the correlator is to find the correlated patterns in noise and to improve the signal-to-noise ratio \cite{Fano1951}. The properties of noise have to be determined to recover the signal. Wiener pioneered in this area in \cite{Wiener1942} and used Volterra expansion for the recovery in the recurrent circuit. These two sides of signal processing remind of the structure of the spiking model of a neural cell \cite{Kistler1997} with plastic Hebbian synapses \cite{Kempter1999}. Indeed, brain information processing might be viewed as a signal processing task. During the signal receiving, the brain aims to make sense of the surrounding environment. Some of the signals represent valuable objects in the form of repetitive patterns. Brain neurons should detect correlations in the input information to recognize such repetitions. For input object detection, neurons have to increase the divergence between reactions to the noise and the regularly appearing objects.

In the present article, we investigate spiking neurons with spike-timing-dependent plasticity (STDP) as active signal recovering devices, seeking correlations in the present input. We propose the modification of the STDP rule \cite{Kempter1999,Song2000} incorporating plasticity substance pool restriction \cite{Nikitin2021}. This modification sets the goal for the neuron to increase the weight divergence between random and correlated input and keep the firing rate of the neuron constrained. The presented model of a neural cell lets neural networks improve the signal-to-noise ratio of a whole network in a layered manner. 

\section{Signal processing and biological groundings for weight divergence optimization approach}

It is remarkable how much in common do modern approaches to signal processing and brain simulation have. However, the connection is often neglected, and a significant legacy of signal processing research seems to us under-looked in theoretical neuroscience. It may include understanding the brain as the optimal filtering device to improve the signal-to-noise ratio to function adaptively in an ever-changing world.

\begin{figure}
\centering
\includegraphics[width=\textwidth]{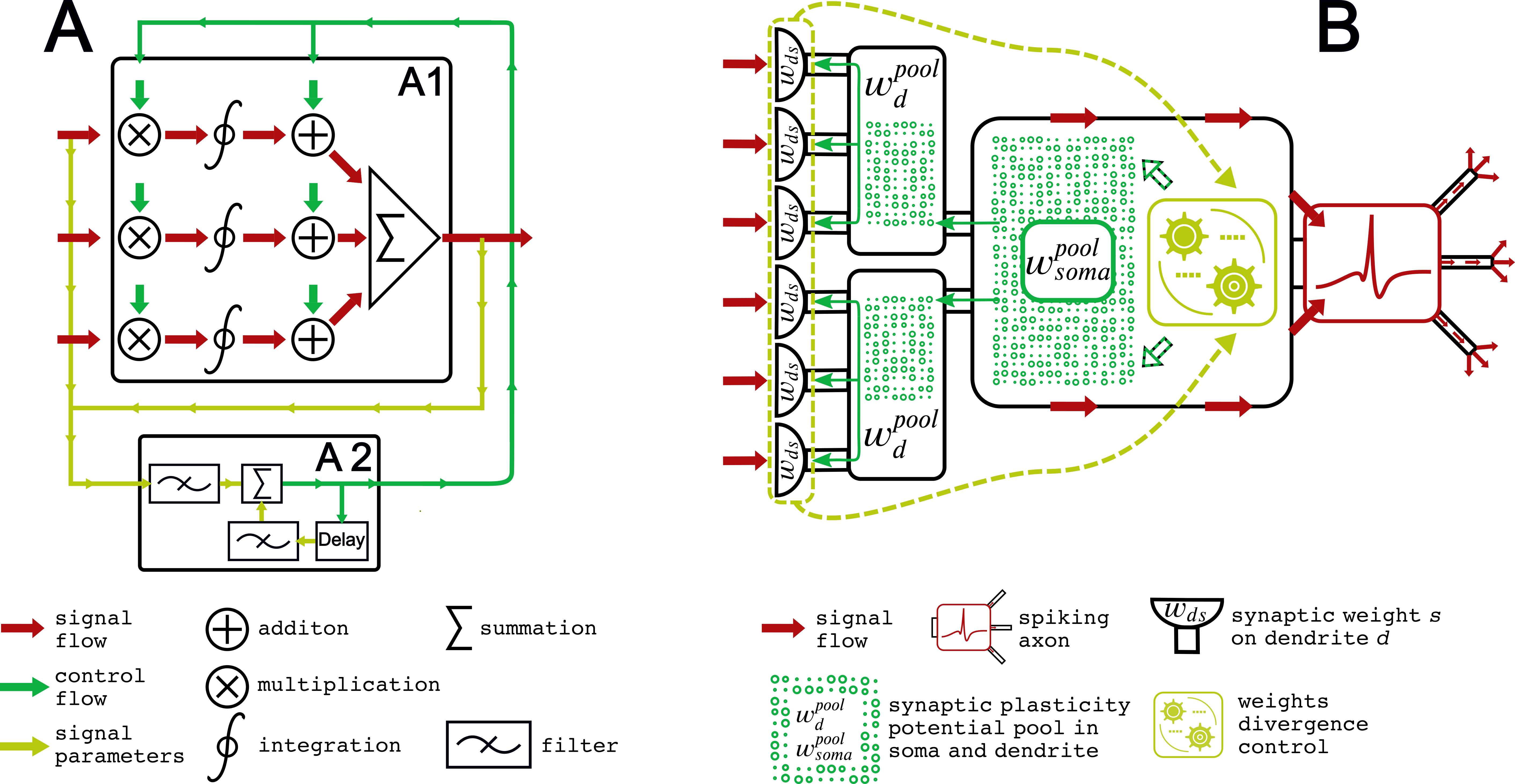}
\caption{The principal schematic of the optimal signal receiver and unsupervised learning neuron. A -- the optimal signal receiver, A1 -- the correlation receiver, A2 -- the noise and signal parameters estimation filter, B -- the spiking neuron with weight divergence maximization control.} 
\label{img:0}  
\end{figure}

A single neuron or small neural assembly task is very similar to some radar stations. The passive radar station constantly scans the surrounding area in search of repetitive patterns, so as the neuron searching for patterns in input spike combinations. For passive signal receivers, it is essential to find correlated, and repetitive unknown patterns in the presence of incoherent noise \cite{Liu2015}. The optimal detector for such a task should incorporate a set of correlation detectors and a nonlinear filter to recognize noise and signal parameters. (see Figure~\ref{img:0})

The most distinctive parts of the filter presented above are synchronized correlators for different signal reception and a recurrent filter to estimate noise and signal parameters. The basic structure of the correlator was proposed by Fano \cite{Fano1951}. The basic idea of correlator lies in adaptive filtration and recurrent reverberation of input. Thus, the goal of the correlator is similar to the Hebb postulate. Indeed, the modern formulation of the STDP principle \cite{Kempter1999,Song2000} raises from the correlation approach to the synaptic weight plasticity in biological neurons. 

The filtering approach for noise parameters estimation was first proposed in \cite{Wiener1942}. However, it is notable that in this paper, Wiener used an approach that later became known as Volterra series expansion and \cite{Kistler1997} also used Volterra expansion for the reduction of the Hodgkin-Huxley model to a more simplified one. Thus, in principle, it shows that cell spiking behavior may, in theory, play the role of noise detector and active coordinator of synaptic plasticity as correlation detector filters.  
 
In our previous work, we have introduced the model of a neuron with STDP synapses and a restricted plasticity pool, used for pro-active control of firing rate \cite{Nikitin2021}. Here, we extend this approach to maximize the noise-to-signal discrimination efficiency of a spiking neuron in an unsupervised reception task. We propose that the goal of many neural cells might be to find and enhance the difference between the correlated repetitive input patterns and background noise. It is consistent with a biologically plausible model \cite{Parise2016} and with earlier proposed research \cite{Livshits2002} and also, with optical hardware-based implementations of well-known neural networks \cite{Sokolov1995}. 

Below, we propose the weight divergence maximization principle as guiding for the restriction of Hebbian plasticity. In this approach, like in the optimal filter proposed above, there is a set of correlators, performing a cross-correlation between different input signals by STDP plasticity restricted by the plasticity potential pool. This pool comprises an abstract substance responsible for synaptic growth and controlled by central cellular processes to improve the correlated signals selectivity. Furthermore, we suppose that some plasticity substance values are spent to let weights grow, restricting the weight amplification. In previous research, we have demonstrated that such an approach allows heterosynaptic plasticity and synaptic competition and amplifies the signal-to-noise ratio of Hebbian plasticity while keeping the firing frequencies of neurons under precise control. Below, we describe the modification of the described approach to amplify the correlated signal transduction in an unsupervised manner.

\section{A weight divergence maximization approach to Hebbian plasticity}

The modification of membrane structures realizes synaptic plasticity in biological neurons at the expense of the secretion of proteins. In the present paper, we will call these weight modification structures the plasticity pool. It consists of abstract substances spending on weight modification. In \cite{Nikitin2021} authors provide the detailed explanation and formulation of dependencies and equations. Here, we will give a brief overview of the basic principle of weight correction restriction. 
In this paper, we consider the Izhikevich \cite{Izhikevich2003} spiking neuron model composed of three dendritic compartments. Dendrites have synaptic inputs with a set of weights ${w}_{ds}$. All signals ${x}_{ds}$ are weighted and summed and transferred to the dynamical excitability model. We suppose that a weight correction substance is secreted in the central compartment of the neuron. It is stored in the soma and called $w^{pool}_{soma}$. The weight plasticity substance is spent from $w^{pool}_{soma}$ to refill the plasticity storage pools $w^{pool}_{d}$ in $d$ dendrites.

The plasticity pool $w^{pool}_{d}$ is then used to provide the resources for synaptic weights amplification. The basic dynamics of weights modification in the paper is equivalent to asymmetric STDP rule \cite{Song2000}. The initial weight correction $\Delta w^{stdp}_{ds}$ is calculated according to the standard STDP rule. We use restriction coefficient of the synaptic weight growth $k^{wp}_{d}$ to determine the final update $\Delta w^{stdp^{**}}_{ds}$ of the synaptic weight of the synapse $s$ on the dendrite $d$:

\begin{equation}\label{eq:1}
\Delta w^{stdp^{**}}_{ds} =
\begin{cases} 
\hfill \Delta w^{stdp}_{ds}, & \text{if } \Delta w^{stdp}_{ds} < 0, \\
k^{wp}_{d} \cdot \Delta w^{stdp}_{ds}, & \text{if } \Delta w^{stdp}_{ds} > 0, \\
\end{cases}
\end{equation}
where $k^{wp}_{d}$ is the restriction coefficient of the growth of synaptic weights depending on the deviation from the optimum frequency.

\begin{equation}\label{eq:2}
k^{wp}_{d} \sim \frac{w^{pool}_{d}}{\sum\limits_{s}{\Delta w^{stdp}_{ds}}}, \text{for all } \Delta w^{stdp}_{ds} > 0,
\end{equation}

According to Equation~\ref{eq:2}, weight growth restriction depends on the control of the $w^{pool}_{d}$ and, consequently, of the $w^{pool}_{d}$ as the initial source of plasticity restricting substance. The neuron is involved in signal separation from noise. We may use the $w^{pool}_{soma}$ level control for the facilitation of correlated input signals separation from noise. In this regard, we calculate $w^{pool}_{soma}$ for the next step depending on the maximum estimated standard deviation $\sigma(w_{ds})$ of synaptic weights.

\begin{equation}\label{eq:3}
\sigma(w_{ds}) = \sqrt{\frac{1}{d\cdot s-1} \sum\limits_{d}\sum\limits_{s}{(w_{ds}-\bar w)^2}},
\end{equation}
where ${w}_{ds}$ -- the current synaptic weight of the synapse $s$ on the dendrite $d$ in time $t$, $\bar w$ -- the mean value of the ${w}_{ds}$ in time $t$. 

We propose an algorithm to maximize the synaptic weight divergence based on $w^{pool}_{soma}$ allocation (see Algorithm~\ref{al:1}). This approach can be called \textbf{weight divergence maximization (WDM)}.

\begin{algorithm}\label{al:1}
\caption{The weight divergence maximization algorithm (WDM) based on $w^{pool}_{soma}$ allocation by the standard deviation of synaptic weights.}
\begin{algorithmic}[1]
\REQUIRE $\Delta w^{stdp}_{ds}$
\STATE Calculate $w^{min}_{ds}$ (if $w^{pool}_{d}=0$) and $w^{max}_{ds}$ (if $w^{pool}_{d}=w^{res}_{d}$)
\STATE Calculate standard deviations: $\sigma(w^{min}_{ds})$ and $\sigma(w^{max}_{ds})$
\STATE Compare $\sigma(w^{min}_{ds})$ and $\sigma(w^{max}_{ds})$:
    \IF{$\sigma(w^{min}_{ds})>\sigma(w^{max}_{ds})$}
        \STATE Set $w^{pool}_{soma}=0$
    \ELSIF{$\sigma(w^{min}_{ds})<\sigma(w^{max}_{ds})$}
        \STATE Set $w^{pool}_{soma}=\sum\limits_{d}\sum\limits_{s}{\Delta w^{stdp}_{ds}}$ $\text{for all } \Delta w^{stdp}_{ds} > 0$
    \ELSIF{$\sigma(w^{min}_{ds})=\sigma(w^{max}_{ds})$}
        \STATE Remain $w^{pool}_{soma}$ the same as in the end of the previous step $t-1$
    \ENDIF
\RETURN $w^{pool}_{soma}$
\end{algorithmic}
\end{algorithm}

Here, $w^{res}_{d}$ is the maximum amplitude of growth of the sum of weights on the dendrite $d$ in $t$ previous steps. We calculate the theoretically possible minimum $w^{min}_{ds}$ (if $w^{pool}_{d}$ is equal to zero) and maximum $w^{max}_{ds}$ (if $w^{pool}_{d}$ is maximum possible) values of the weights. Next, we determine the standard deviation from the resulting values ($\sigma(w^{min}_{ds})$ and $\sigma(w^{max}_{ds})$) and compare them. Based on this comparison, we set the $w^{pool}_{soma}$, which should lead to maximization of the weights divergence for the current step.

\subsection{Firing rate-controlling approaches to WDM}

The basic WDM algorithm should maximize the difference between informative and random inputs. However, it set no goal for the neuron firing rate, and it will fluctuate according to input signal parameters. The approach above seems to be plausible but not flexible enough. We cannot set it to some determined goal point. At the same time, for some tasks such as switching between different network oscillation frequencies, it is crucial to keep the target frequency. That is why it is necessary to drive the current firing rate $\theta_{real}$ to some set point as target firing rate $\theta_{target}$. We calculate the difference ($\Delta \theta$) between $\theta_{target}$ and $\theta_{real}$ to achieve the selected goal. If the firing rate is too low, we satisfy all the demand for plasticity modification ($\sum\limits_{d}\sum\limits_{s}{\Delta w^{stdp}_{ds}}$) and essentially act for the next step as general STDP. In the case when the firing rate $\theta_{real}$ is too high $w^{pool}_{soma}$ is set to zero. Actions in both deviation cases are done with some probability $p(\Delta \theta)$, proportional to the current deviation. We call this approach a \textbf{firing rate optimization (FRO)} of the WDM (see Algorithm~\ref{al:2}).

\begin{algorithm}\label{al:2}
\caption{The weight divergence maximization algorithm with firing rate-controlling.}
\begin{algorithmic}[1]
\REQUIRE $\theta_{real}$, $\theta_{target}$
\STATE Perform all the steps of the Algorithm~\ref{al:1}
\STATE Calculate $\Delta \theta = \theta_{real} - \theta_{target}$
    \IF{$\Delta \theta>0$}
        \STATE Calculate the probability $p(\Delta \theta) = \beta^+ \cdot \Delta \theta$
        \STATE With $p(\Delta \theta)$ probability, set $w^{pool}_{soma} = 0$
    \ELSIF{$\Delta \theta<0$}
        \STATE Calculate the probability $p(\Delta \theta) = - \beta^- \cdot \Delta \theta$
        \STATE With $p(\Delta \theta)$ probability, set $w^{pool}_{soma}=\sum\limits_{d}\sum\limits_{s}{\Delta w^{stdp}_{ds}}$ $\text{for all } \Delta w^{stdp}_{ds} > 0$
    \ENDIF
\RETURN $w^{pool}_{soma}$
\end{algorithmic}
\end{algorithm}

A firing rate minimization represents the special case. Indeed, for the biological brain, it is crucial to minimize the energy spent on spiking. Hence, it is most apparent to minimize the firing rate along with STD maximization. We call it a \textbf{firing rate minimization (FRM)} of the WDM approach. In this case, the $\theta_{target}$ is set to zero, meaning that the frequency will always be minimal, while weights' standard deviation (SD) will be maximized. We introduce the amplification coefficients $\beta^-$/$\beta^+$ to control the sensitivity to minimization/maximization of firing rate and apply them to $p(\Delta \theta)$. In our experiments below, the $\beta^+$ and $\beta^-$ equal to one.

\section{Different noise filtering tasks examination}

In order to test the possibilities of using the methods described above, we propose to carry out tests with various combinations of noise and input signals. For each synapse, the signal in the experiments is a time series of binary data, where the numbers of "1" appear at random with a specific predetermined frequency. Since the data is binary, we calculate the closeness of the correlation for correlated signals between different synapses using the Matthews correlation coefficient. Further, we compare the proposed models with the model of the basic STDP and \textbf{the model of backward calculation of plasticity potential reserve demand (PPD)} proposed in \cite{Nikitin2021}. The neuronal plasticity regulation model changes the potential synaptic weight growth by controlling the plasticity potential pool to stabilize the artificial neuron's firing rate dynamically.

Experiments are carried out on the neuron with three dendrites, each of which has six synapses. We run 100 tests with 2400 steps. The results for the figures are averaged. 

\subsection{Detection of signals with different correlation coefficients}

In the following experiment, the nature of the signal incoming to the neuron varied from noise to a correlated signal. We send correlated input to the chosen synapses during specific time ranges. The signal frequency for all synapses is 0.2 throughout the experiment. The correlation coefficient for chosen synapses from 100 to 300 steps and from 1300 to 1500 steps is 0.5, from 500 to 700 steps and from 1700 to 1900 steps is 0.7, and from 900 to 1100 steps and from 2100 to 2300 steps is 0.9. The rest of the synapses during these periods receive noise with a frequency of 0.2.

\begin{figure}
\centering
\includegraphics[width=\textwidth]{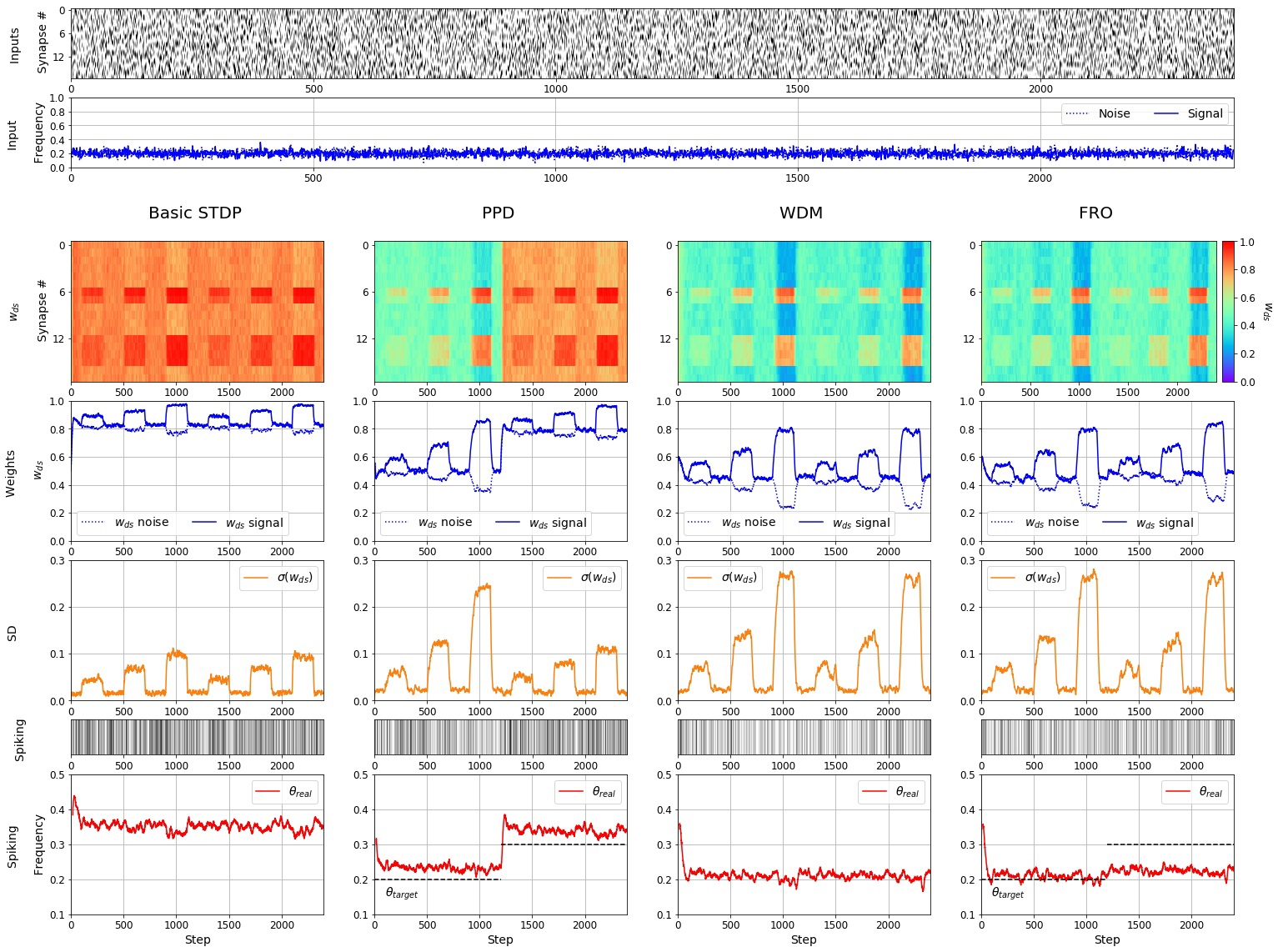}
\caption{Detection of signals with different correlation coefficients.} 
\label{img:1}  
\end{figure}

In Figure~\ref{img:1}, you can see that all models, including STDP, are capable of filtering noise by extracting correlated signals. The higher the correlation coefficient between the incoming signals, the better the correlated signal differs from the noise: with a low correlation coefficient, the SD value is significantly lower than with a high correlation coefficient, and vice versa. It can be seen that the basic STDP copes with this task worse than other models and also adheres to a high current firing rate in comparison with other models due to an increase in synapse weights. Increasing the target firing rate by 0.1 for the PPD model degrades its results significantly, making them close to the baseline PPD. It is due to the minimization of the $w^{pool}_{soma}$ deficit, which removes the restriction on the growth of synapse weights. The same increase in the target firing rate for the FRO model results in a slight degradation in quality and a slight increase in the firing rate. The WDM model independently goes into a state with a current firing rate close to the frequency of the incoming signals. We can see that all the weight-restricted approaches (PPD, WDM, FRO) improve the signal-to-noise ratio relative to the standard STDP plasticity rule.

\subsection{Detection of a correlated signal from low-frequency noise}

In the second experiment, we input the low-frequency noise and change the signal parameters for the chosen synapses. 

\begin{figure}
\centering
\includegraphics[width=\textwidth]{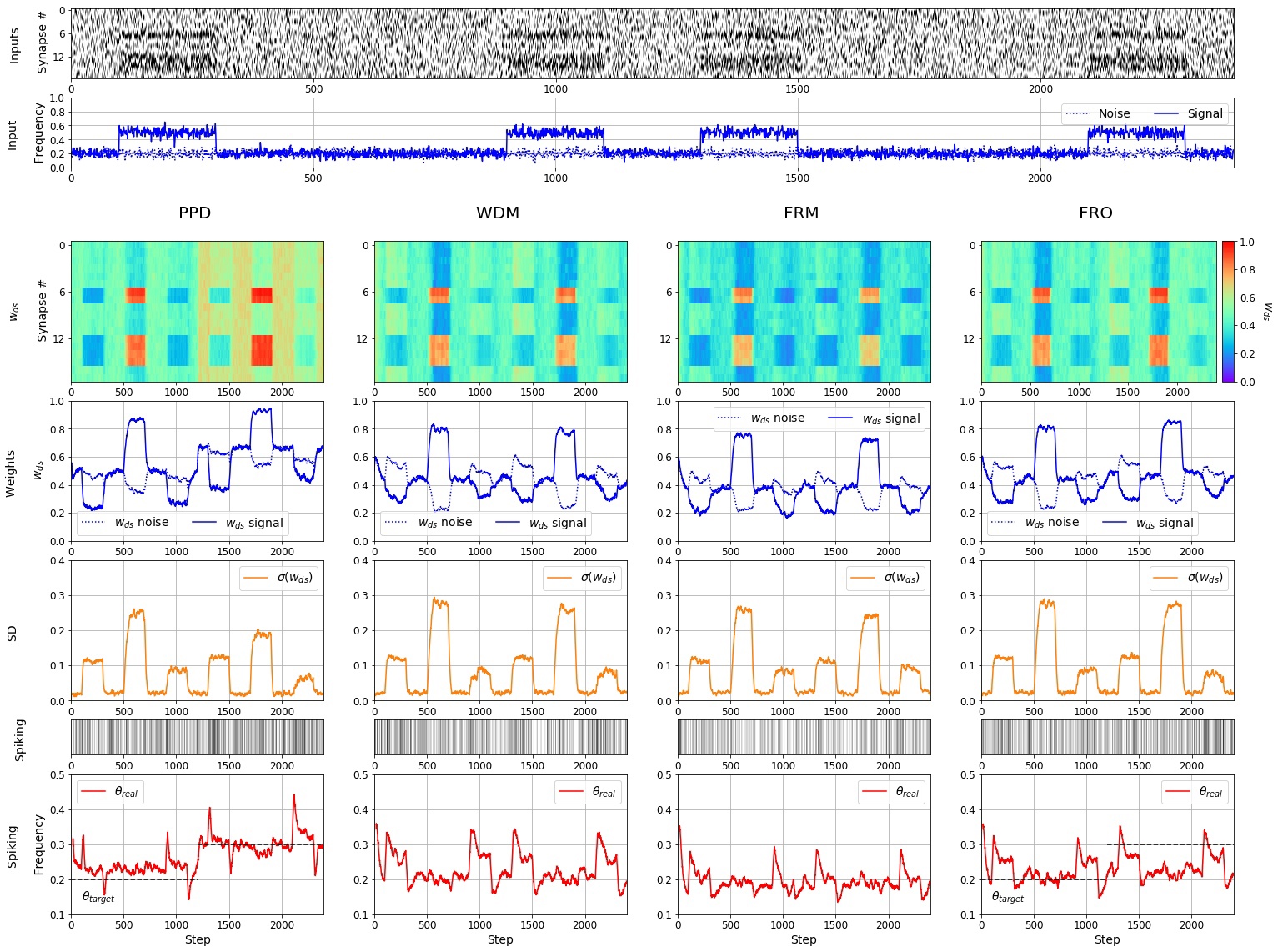}
\caption{Detection of the correlated signal from low-frequency noise with low target firing rate. Here, the input frequency is constant, and the correlation between input signals varies over time.} 
\label{img:21}  
\end{figure}

So, we apply the correlated signals to 2 synapses of one dendrite and 4 synapses of another. A signal of different frequencies arrives at these synapses with or without correlation in specified ranges. All other synapses always receive the noise.

We set the ranges for chosen synapses as follows: from 100 to 300 steps and from 1300 to 1500 steps, we input the noise with a frequency of 0.5, from 500 to 700 steps, and from 1700 to 1900 steps, synapses receive a correlated signal with a frequency of 0.2, from 900 to 1100 steps and from 2100 to 2300 steps a correlated signal with a frequency of 0.5 is received (high-frequency correlated input). The rest of the synapses during these periods receive noise with a frequency of 0.2. In all remaining step ranges, all synapses of the neuron receive noise with a frequency of 0.2.

The plot of synapse weights in Figure~\ref{img:21} shows that all models filter out high-frequency noise and high-frequency correlated signals entering the neuron simultaneously with low-frequency noise. 

\begin{figure}
\centering
\includegraphics[width=\textwidth]{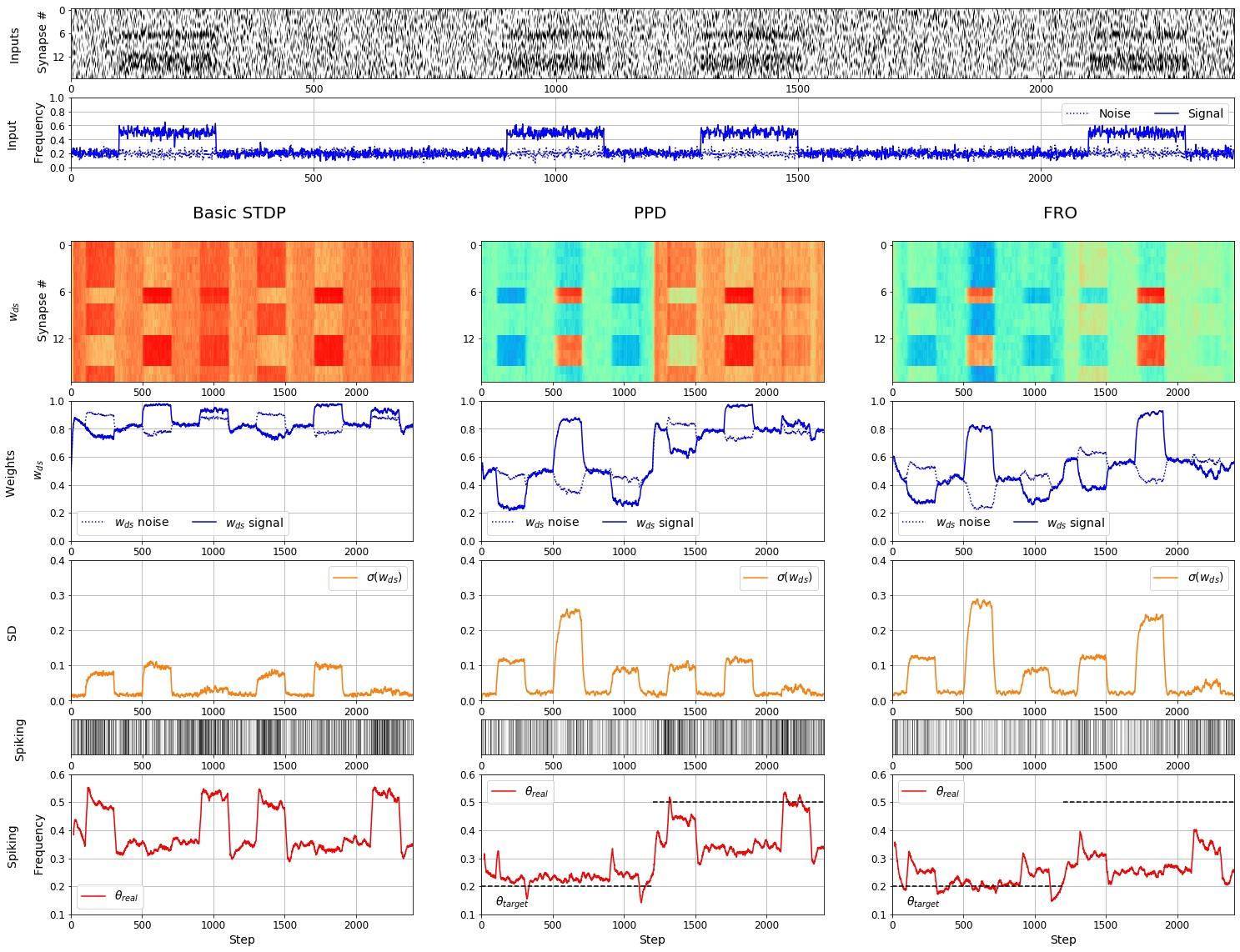}
\caption{Detection of a correlated signal from low-frequency noise with high target firing rate.} 
\label{img:22}  
\end{figure}

At the same time, the plots of the SD show that the FRM model manages to achieve slightly higher SD values on the ranges with high-frequency correlated input. Also, in these areas, this model shows the lowest firing rate. All weight divergence models operate at approximately the same firing rate during the periods of low-frequency correlated input. At the same time, the model aimed at maximizing weight divergence reaches the highest SD values. Setting the target firing rate to 0.3 worsens the test results for the PPD model but almost does not change the results of the FRO model.

Figure~\ref{img:22} shows the results of modeling the same task, but for the basic STDP model, for PPD and FRO with a target firing rate equal to 0.5. Again, we see that only basic STDP and PPD with a high target firing rate can filter high-frequency correlated signals from low-frequency noise. In the case of PPD, we achieve it through a significant reduction in the $w^{pool}_{soma}$ deficit, which brings this model closer to the basic STDP. At the same time, PPD effectively separates high-frequency noise from low-frequency noise. In the case of FRO, when the target firing rate increase, the model still effectively separates the low-frequency correlated signal from the low-frequency noise but continues to filter out the high-frequency correlated signal, albeit to a lesser extent than with a low target firing rate.

\begin{figure}
\centering
\includegraphics[width=\textwidth]{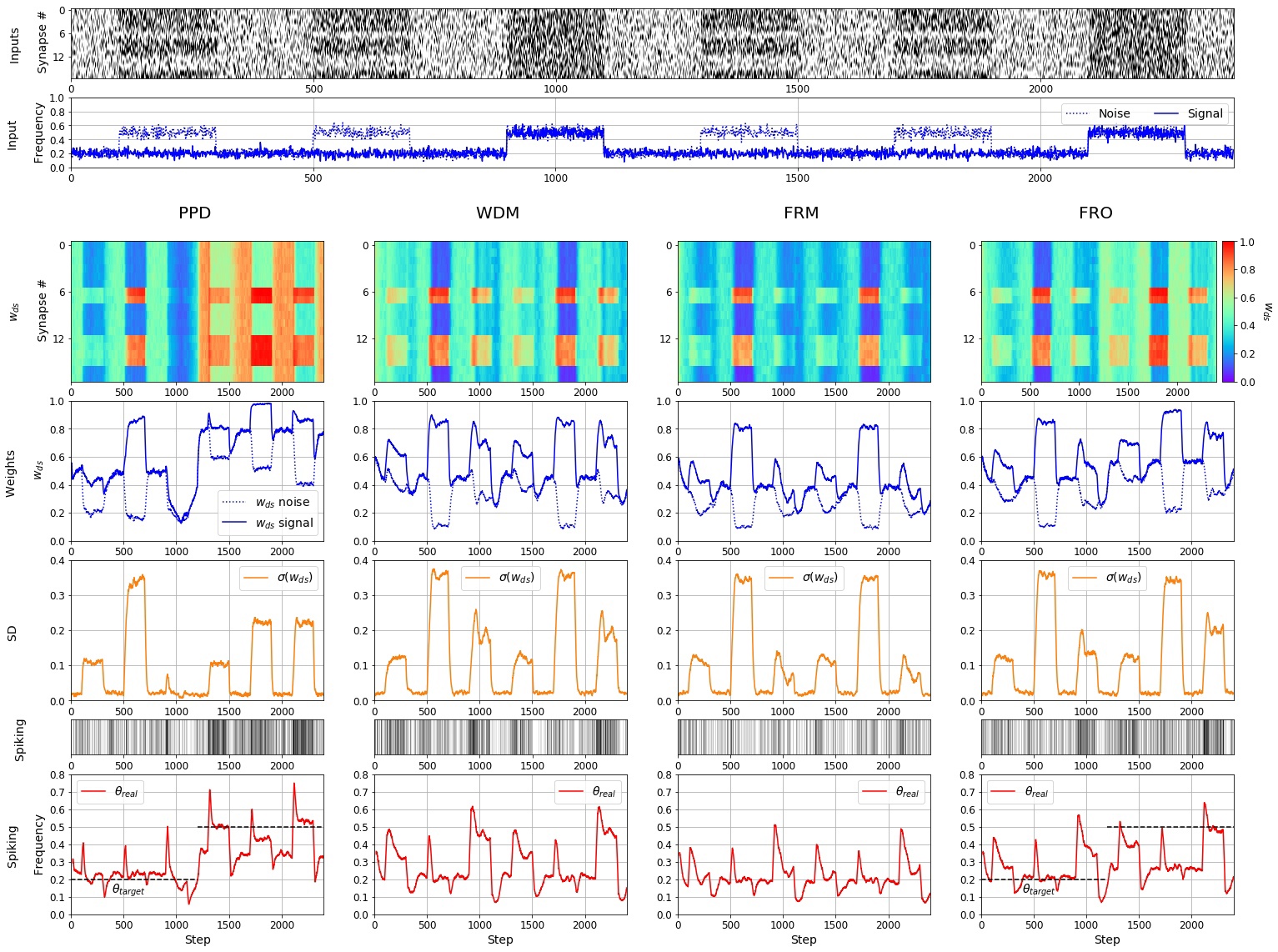}
\caption{Detection of a correlated signal from high-frequency noise.} 
\label{img:3}  
\end{figure}

\subsection{Detection of correlated signal from high-frequency noise}

In this experiment, we assess the change in signal parameters for chosen synapses using high-frequency noise. Thus, we apply the correlated signals to 2 synapses of one dendrite and 4 synapses of another. A signal of different frequencies arrives at these synapses with or without correlation in specified ranges. All other synapses always receive the noise. These parameters are the same as in the first experiment. In the difference with the last part, we present the functioning of models in the presence of more intense uncorrelated high-frequency noise. The ranges for the chosen synapses in this task are as follows: from 100 to 300 steps and from 1300 to 1500 steps, we input the noise with a frequency of 0.2, from 500 to 700 steps, and from 1700 to 1900 steps, synapses receive a correlated signal with a frequency of 0.2, from 900 to 1100 steps and from 2100 to 2300 steps we input a correlated signal with a frequency of 0.5. The rest of the synapses during these periods receive noise with a frequency of 0.5. In all remaining step ranges, all synapses of the neuron receive noise with a frequency of 0.2.

Figure~\ref{img:3} shows that in areas where low-frequency and low-frequency noise are combined, the PPD model reduces the weights at synapses with high-frequency noise, filtering them out. At the same time, the WDM model increases the weights of synapses with low-frequency noise. The FRM and FRO models simultaneously increase the weights of synapses with low-frequency noise and filter out synapses with high-frequency noise. Thus, all four models achieve approximately the same SD value in these areas. It is also unaffected by setting a high spike rate target. During the periods of low-frequency correlated input, all four models achieve high SD values. Setting a high target firing rate for the PPD model degrades the simulation results significantly. However, for the FRO model, the SD indicator decreases insignificantly. In the area with a high-frequency correlated input and a low target firing rate, the PPD model equally reduces the weights of all neuron synapses, filtering out the input signals for them. With the nature of the inputs, the PPD model can successfully filters high-frequency noise only when setting high target values of the firing rate. The FRM model also copes with this task worse than the WDM model, reducing its ability to filter out high-frequency noise from high-frequency correlated signals due to its tendency to reduce the firing rate after their sharp increase. Although it reaches various SD values, the FRO model successfully copes with this task both at high and low target values of the firing rate.

Thus, the models discussed above have both advantages and disadvantages, depending on the specific task. For example, the WDM model shows itself better than other models in detecting a low-frequency correlated signal, but it increases the firing rate when working with high-frequency signals. On the other hand, the FRM model allows one to restrain the growth of the firing rate, but this ability degrades the performance when high-frequency signals with high-frequency noise arrive. At the same time, the FRO model allows one to optimize the firing rate without significant losses in the quality of work. However, its practical application involves the setting of a target firing rate.

\section{Discussion and perspectives} 

In the present paper, we have drawn the connections between signal reception approaches and the biological basement of neural activities. We have proposed an unsupervised approach to noise filtering and signal reception based on Hebbian plasticity. This approach is biologically plausible and based on physical restrictions of the synaptic modifications of real biological neurons. Examining the performance of restricted STDP in the presence of different signals and noise intensities showed that the physically bounding of STDP plasticity and goal-directed control of the synaptic growth led to significant improvement of the signal-to-noise ratio of a single neuron as a signal receiver. The plasticity rules in the paper are unsupervised and local, as well as the standard STDP rule. The rule showed in the present paper was studied for one stimulated neuron and needs to be studied in networks, including networks consisting of excitatory and inhibitory neurons. The frequency control model (FRO) will stabilize the overall frequencies in the whole neural net, allow switching of neural assemblies in different oscillation regimes, and allow replay of different memorized patterns inside the minimal neural network circuit.

\section*{Acknowledgement}
The computing resources of the Shared Facility Center "Data Center of FEB RAS" (Khabarovsk) were used to carry out calculations.

\bibliographystyle{unsrt}
\bibliography{mybibliography}

\end{document}